\documentclass[final,3p,times,twocolumn]{elsarticle}
\usepackage{graphicx}
\usepackage{amssymb}
\usepackage{amsmath}
\usepackage{hyperref}
\usepackage{color}
\usepackage{gensymb}

\journal{Commun. Nonlinear Sci. Numer. Simul.}
\begin{document}

\title{Deep inelastic collision of two-dimensional anisotropic dipolar condensate solitons}

\author[fac]{Luis E. Young-S.}
\ead{lyoung@unicartagena.edu.co}
\cortext[author]{Corresponding author.}
 \address[fac]{Facultad de Ciencias Exactas y Naturales, Universidad de Cartagena, 130014 Cartagena, Bolivar, Colombia}
\author[int]{S. K. Adhikari\corref{author}}
\ead{sk.adhikari@unesp.br}
\ead[url]{professores.ift.unesp.br/sk.adhikari/}
 \address[int]{Instituto de F\'{\i}sica Te\'orica, UNESP - Universidade Estadual Paulista, 01.140-070 S\~ao Paulo, S\~ao Paulo, Brazil}

\begin{abstract}
The possibility of generating stable anisotropic solitons in dipolar Bose-Einstein condensates (BECs) leads  to a different scenario  not possible in a nondipolar BEC with   contact interaction.
We study the  statics and dynamics of anisotropic bright solitons in quasi-two-dimensional  BECs
consisting of polarized dipolar atoms.
We study the collision dynamics of two such solitons at different velocities
for different angles between the polarization and collision directions.
The collision is found to be  quasi elastic  at large velocities. At small velocities 
the collision is  inelastic leading to the formation of a coalesced soliton in an excited 
scissors mode, monopole mode or quadrupole mode. Also, at small velocities, after collision, 
a large change of direction of motion of the solitons is possible. The investigation is performed 
through a numerical solution of the underlying mean-field Gross-Pitaevskii equation.

\end{abstract}
 

\maketitle

\section{Introduction}

A bright soliton \cite{sol} is a self-bound localized object, 
that maintains its shape, while moving at a constant velocity 
in one-dimension (1D),  due to a cancellation of dispersive effects and nonlinear attraction. 
The solitons can be both classical and quantum mechanical in nature. 
 Quasi-one-dimensional (quasi-1D) solitons  and soliton trains were created { in a laboratory by considering strong traps in two transverse directions}
and studied  experimentally in  nondipolar quantum  Bose-Einstein condensate (BEC) of $^7$Li \cite{417N02}
and $^{85}$Rb atoms \cite{solrb} following a theoretical suggestion \cite{th} in addition to 1D solitons 
in classical systems like  water waves and 
nonlinear optics \cite{sol}. 
The quasi-1D geometry is created in a BEC by subjecting it to  strong  transverse traps along 
$x$ and $y$ directions and allowing the soliton  to move freely along the $z$ direction.  
  The  BEC 
solitons constitute a fascinating consequence of a balance between the  quantum-mechanical nonlinear attraction  and the   quantum-mechanical dispersive 
pressure thus  resulting in self-confined wave-packets.  {The study of quasi-1D BEC solitons continues  to be a very active field of research to date \cite{hulet}.}
However,  the   nondipolar BEC solitons  are fundamentally unstable in two \cite{2d} and three \cite{sol,3d1} dimensions leading either to collapse or {dispersion}  \cite{nova}.

After the experimental observation of BECs of polarized  $^{52}$Cr \cite{beccr}, $^{164}$Dy \cite{becdy}, and $^{168}$Er \cite{becer}
atoms with large magnetic dipole moment, there has been a rapid growth of  interest in dipolar BECs \cite{dipbec}, 
which differ significantly from their non-dipolar counterparts due to the anisotropic 
long-range character of the dipole-dipole interaction. Polar molecules with large electric dipole moment were also 
considered for BEC experiments \cite{polar}. 
In addition to the usual quasi-1D nondipolar and dipolar BEC
solitons, one can also  have quasi-two-dimensional (quasi-2D) { isotropic  \cite{2dsol1} as well as anisotropic
 \cite{2dsol2}   solitons
in dipolar BECs polarized along the $z$ direction.  
A quasi-2D setting can be obtained in a  laboratory 
by considering a strong trap in one direction.
For a quasi-2D  isotropic soliton the strong trap is in the polarization direction and for a quasi-2D  anisotropic soliton the strong trap is in a direction perpendicular to the polarization direction.}
 Moreover, solitons in a dipolar BEC remain stable when
the transverse harmonic trap is  replaced by a periodic optical-lattice trap in  the 
 quasi-2D \cite{17} configuration. 
Both ground and exited states of 
quasi-2D   solitons  are { found to be}  stable in polarized
dipolar BECs under proper values of nonlinear  dipolar and contact atomic interactions. 
More specifically,
 these quasi-2D dipolar solitons {could be  stable} 
for a large repulsive nonlinear strength of contact repulsion 
with an even larger nonlinear strength of dipolar interaction \cite{2dsol1,2dsol2,102PRL09}, 
e.g., for a large atomic scattering length $a$ and for $a_{\mathrm{dd}}>a$, where $a_{\mathrm{dd}}$ is a dipolar length, viz. Eq. (\ref{eq.dl}), 
to measure the strength of dipolar interaction.  {Nevertheless, in general, an excited state can also be unstable. }
A large value of $a$ corresponding to a large short-range repulsion
inhibits collapse to a great extent, whereas a large $a_{\mathrm{dd}}$ leads to a large long-range attraction to bind the soliton. 
Although, the dipolar interaction is repulsive in the direction perpendicular to polarization $z$, a harmonic trap in the perpendicular direction $y$ suffices to form a quasi-2D robust {anisotropic} dipolar soliton  free to move in the $x-z$ plane \cite{2dsol2}. 
In spite of being one of the major novel features of the nonlinear physics of dipolar BEC, these quasi-2D solitons 
have not yet been observed in experiments, {although  quasi-2D Townes solitons \cite{2d} in a BEC were observed in two recent experiments \cite{2dsolx}.}
{ In a different setting in nonlinear optics, however, a stable quasi-2D spatial soliton has been observed in liquid carbon disulfide \cite{cid}.}

Because of these interesting possibilities in dipolar
BEC, we study here some  novel features of the dynamics of    {anisotropic}  quasi-2D  solitons \cite{2dsol2} in a polarized dipolar BEC.
Specifically, we investigate the collision dynamics of two robust  quasi-2D dipolar solitons 
in the $x-z$ plane by varying the initial relative velocity and initial angle of impact using a mean-field Gross-Pitaevskii (GP) equation.  
In particular, we use two quasi-2D solitons  with different angles between the  collision 
directions of the two solitons  in order to study a remarkable novel nonlinear feature of dipolar dynamics. 
At large  velocities
 the dynamics is quasi-elastic for both {head-on collision and collision with a non-zero impact parameter}
and the solitons pass through each other 
and move in straight line without significant deformation in shape. 
{As the velocity is reduced the collision becomes inelastic and the emerging solitons become deformed.
With further reduction in velocity the two solitons form a coalesced soliton in an excited state which eventually breaks up into { few pieces that come out} in directions independent of the initial directions of motion before collision.}  
At very small initial velocities, the solitons come close to each other and form a  coalesced single soliton  and never separate. { These latter inelastic processes through the formation of a coalesced soliton at small velocities will be  called {\it deep inelastic collision}.}
These  coalesced solitons are excited states of solitons where the initial kinetic energy is partially 
transformed into 
the internal excitation energy of the final soliton. 
{In deep inelastic collision, after formation, the coalesced soliton often executes {internal
oscillations}. 
The simplest mode of oscillation is monopole or breathing mode (radially symmetric contraction and expansion maintaining the shape.) The next excitation modes are dipole mode (contraction and expansion along one direction with change of shape) and   quadrupole mode (simultaneous contraction and expansion along two directions with change of shape.)}
We find different types of excitation of the coalesced soliton. Specifically, we find a signature of 
scissors-mode and monopole mode  
oscillations \cite{sciex} and  dipole and  quadrupole excitations in the  coalesced soliton. 
{ Similar {studies} of collision of two quasi-1D dipolar solitons have identified the formation of a breathing mode \cite{bm}.}
{ The 
free} angular oscillation of a BEC around an axis is called a scissors-mode oscillation.  
The scissors-mode oscillation in a BEC  
was detected \cite{sciex} in the  oscillation of a $^{87}$Rb BEC  excited by a sudden rotation of the anisotropic trapping potential.
In a nondipolar BEC, 
scissors-mode oscillation was initiated 
by external rotation \cite{sciex}; here scissors-mode oscillation is initiated by inelastic collision between two dipolar solitons, while a part of the initial kinetic energy is transformed into angular  oscillation  energy. { In case of scissors-mode oscillation the oscillation eventually die out and the solitons form a fused soliton in an excited state (soliton fusion).}
Also, at small 
initial velocities, a special coupling with internal soliton modes can appears and, interestingly enough, 
the direction of propagation of the solitons may change after collision. In numerical simulation we find  
a change of direction of propagation  after collision.  {Many of the issues found in the numerical simulation of collision of two quasi-2D anisotropic dipolar solitons,  including quasi-elastic
reflection
at large velocities and capture at small velocities, are encountered in other
settings
including collision between kink and antikink traveling waves of the $\phi^4$ equations
of non-integrable field theories \cite{siam}.    Formation of coalesced soliton molecules has also been studied experimentally in nonlinear fiber optics \cite{fiber}.}

In Sec.    \ref{sec2} we describe the  three-dimensional (3D) mean-field GP equation for a dipolar BEC. 
A reduced quasi-2D GP equation, which we employ for the present study of quasi-2D dipolar solitons, is given in Sec. \ref{2b}.  In Sec. \ref{sec3} we describe the numerical results for a stationary quasi-2D soliton.  Those for the  dynamics of collision between two quasi-2D solitons at different velocities and for different angles of impact are presented in Sec. \ref{sec4}. Specifically, in Sec. \ref{4.1} we consider quasi-elastic
head-on  collision at large velocities.
In Sec. \ref{4.2} quasi-elastic collision with a non-zero impact parameter is considered.
 In Sec. \ref{4.3} results for deep inelastic collision at small velocities are presented; in Sec. \ref{4.2.1} we consider  a scissors-mode oscillation, in Sec. \ref{4.2.2}  an unexpected change in the direction of the emerging solitons, and finally in Sec.\ref{4.2.3}  a quadrupole excitation of the coalesced soliton are considered.
 Finally,  in Sec. \ref{sec5} we provide a summary of our findings and conclusions.

\section{Mean-field model for anisotropic dipolar BEC soliton}

\subsection{3D Gross-Pitaevskii equation}
\label{sec2}
We consider a BEC of $N$ polarized dipolar atoms of mass $m$ each,
interacting via the following 
 atomic dipolar and contact interactions   \cite{dipbec,dip}
\begin{eqnarray}
V({\bf R})= 
\frac{\mu_0 \mu^2}{4\pi}\frac{1-3\cos^2 \theta}{|{\bf R}|^3}
+\frac{4\pi \hbar^2 a}{m}\delta({\bf R }),
\label{eq.con_dipInter} 
\end{eqnarray}
where $a$ is the atomic scattering length,  $\mu$ is the permanent magnetic dipole moment,
$\mu_0$ is the permeability of vacuum, 
${\bf R}= {\bf r} -{\bf r}'$ is the vector joining two dipoles placed at $\bf r$ ($\equiv \{x,y,z\}$) and $\bf r'$
and $\theta$ is the angle made by ${\bf R}$ with the polarization 
$z$ direction.
At sufficiently low temperatures a dipolar BEC is described by the following  nonlocal 3D GP equation \cite{dipbec,dip}
\begin{align}&
i \hbar \frac{\partial \psi({\bf r},t)}{\partial t}=
{\Big [}  -\frac{\hbar^2}{2m}\nabla^2
+U_{\mathrm{ext}}({\bf r})
+ \frac{4\pi \hbar^2}{m}{a} N \vert \psi({\bf r},t) \vert^2
\nonumber\\ &
+ N \frac{ \mu_0 \mu^2 }{4\pi }
\int U_{\mathrm{dd}}({\mathbf R})\vert\psi({\mathbf r'},t)\vert^2 d{\mathbf r}'\Big] \psi({\bf r},t),
\label{eq.GP3d}
\end{align}
where $U_{\mathrm{ext}}$ is the external trap,
the wave function is subject to normalization $\int \vert \psi({\bf r},t) \vert^2 d{\bf r}=1$
and \mbox{$U_{\mathrm{dd}}({\bf R})=(1-3\cos^2 \theta)/|{\bf R}|^3$}.

In this study of quasi-2D solitons we consider a harmonic trap $U_{\mathrm{ext}}=\textstyle  \frac{1}{2}m\omega_y^2y^2$  along the transverse $y$ direction, where $\omega_y$ is the trap frequency.
To compare the dipolar and contact interactions, the strengths of the dipolar 
interaction is  expressed in terms of the dipolar length
\begin{align}
a_{\mathrm{dd}}=\frac{\mu_0 \mu^2 m }{ 12\pi \hbar ^2}.
 \label{eq.dl}
 \end{align}
The dimensionless ratio  $a_{\mathrm{dd}}/a$ characterizes
the strength of dipole-dipole interaction $a_{\mathrm{dd}}$  compared to that of  the short-range interaction $a$.

Equation (\ref{eq.GP3d}) is next expressed 
in the following dimensionless form 
\begin{align}
i\frac{\partial \psi({\bf r},t)}{\partial t}&={\Big [}  
- \frac{\nabla^2}{2}
+\frac{y^2}{2}  
+ g\vert \psi ({\bf r},t)\vert^2 
\nonumber \\  &  \,
+  3 a_{\mathrm{dd}} N
\int U_{\mathrm{dd}}({\mathbf R})\vert\psi({\mathbf r'},t)
\vert^2 d{\mathbf r}'
{\Big ]}  \psi({\bf r},t),
\label{eq.GP3dHO}
\end{align}
where $g=4\pi Na$.
In Eq. (\ref{eq.GP3dHO}), length is expressed in units of 
oscillator length  $l_0=\sqrt{\hbar/(m\omega_y)}$, 
energy in units of oscillator energy  $\hbar\omega_y$, density 
$|\psi|^2$ in units of $l_0^{-3}$, and time in units of $ 
t_0=1/\omega_y$.

\subsection{Quasi-2D Reduction}
\label{2b}

In the presence of a harmonic potential along the $y$-direction, 
it is natural to assume
that the dynamics of the BEC in the $y$ direction
is confined in the ground state $\phi(y)$ of the harmonic potential
and we have the following ansatz of the wave function 
\begin{align}
\psi({\bf r},t)\equiv \phi(y) \times \phi(\boldsymbol \rho{},t)
=\frac{1}{(\pi l_0^2)^{1/4}}\exp\left[-\frac{y^2}{2l_0^2}  
\right] \phi(\boldsymbol \rho{},t),
\label{eq.an2xz}
\end{align}
where $\boldsymbol \rho{}\equiv \{x,z\}$, and 
$\phi(\boldsymbol \rho{},t)$ is the asymmetric 
effective 2D wave function.
To derive the effective 2D equation for the anisotropic  quasi-2D dipolar BEC,
we use  ansatz (\ref{eq.an2xz}) in Eq. (\ref{eq.GP3dHO}), 
multiply it  by the ground-state wave function $\phi(y)$ and integrate over $y$ to get \cite{2dsol2,dip,luca}:
\begin{align}&
i\frac{\partial \phi(\boldsymbol \rho,t)}{\partial t}
={\Big [} 
-\frac{\nabla_\rho^2}{2}
+\frac{g}{\sqrt{2\pi}l_0}\vert  \phi (\boldsymbol \rho,t) \vert  ^2
\nonumber \\  &  \,
+ \frac{ 3 a_{\mathrm{dd}} N}{\sqrt{2\pi}l_0}
\int d \boldsymbol \rho' \
U_{\mathrm{dd}}^{2D}(\boldsymbol \rho -\boldsymbol \rho') 
|\phi(\boldsymbol \rho{}', t)|^2
{\Big ]}   \phi(\boldsymbol \rho,t).
\label{eq.GP2d}
\end{align}
The dipolar integral is calculated in momentum space by the following convolution rule \cite{goral}
\begin{align}
\int d \boldsymbol \rho' \ & U_{\mathrm{dd}}^{2D}(\boldsymbol \rho -\boldsymbol \rho') 
|\phi(\boldsymbol \rho{}', t)|^2 \equiv 
 \nonumber \\  &  \,
\frac{4\pi}{3}\int \frac{d{\bf k}_\rho}{(2\pi)^2}
 e^{-i{\bf k}_\rho.\boldsymbol \rho} \
 \widetilde n({\bf k}_\rho,t) \
 h_{2D} \biggr(\frac{k_\rho l_0}{\sqrt 2}\biggr),
\end{align}
where $k_\rho=\sqrt{k_z^2+k_x^2}$,
$\widetilde n$ the Fourier transformation of the 2D density \cite{2dsol2,bohn}
\begin{equation}
\widetilde n({\bf k}_\rho,t)=\int d{\bf \rho}e^{i{\bf k}_\rho \cdot {\boldsymbol \rho}}
|\phi({\boldsymbol \rho},t)|^2,
\end{equation}
and  
\begin{eqnarray}
h_{2D}(\xi)  &\equiv & \frac{1}{2\pi}\int^\infty_{-\infty} dk_y \left[  \frac{3k_z^2}{{\bf k}^2}  -1 \right] |{\;}\widetilde n(k_y)|^2,
 \nonumber \\  
&=&  -1+3\sqrt \pi \frac{k_z^2 l_0^2}{2\xi} \exp(\xi^2)\ \text{erfc}(\xi),
\end{eqnarray}
where $ \xi=k_\rho l_0/\sqrt{2}$  and $ \text{erfc}(\xi)$ is the complementary error function.

{ In two dimensions, { in free space, }
a stable nondipolar BEC soliton cannot be formed due to a collapse instability \cite{2d}. 
For small values of nonlinear attraction such an object will escape to infinity and for large values of nonlinear attraction it will collapse. 
However, a weakly attractive BEC in a quasi-2D trap, below the Townes limit \cite{2d} of collapse, can be bound as the escape to infinity will not materialize due to the trap. } { It has been demonstrated \cite{2dsol2} that, for moderate values of the scattering 
length $a$ and of the strength of dipolar interaction, the quasi-2D dipolar equation (\ref{eq.GP2d}) without any external trap
 permits the formation of  a spatially anisotropic soliton which can freely move in the $x-z$ plane.
This soliton is supported only by a trap in the $y$ direction. Equation (\ref{eq.GP2d}) is independent of the variable $y$ as the dependence on $y$ has been integrated out.
Such a soliton will be extended in the $x-z$ plane   and is, in general, called a quasi-2D soliton.
Nevertheless, such a soliton collapses for large values of the strength of dipolar interaction (strong attraction for $a_{\mathrm{dd}}/a \gg 1$)
and escapes to infinity for large values of scattering length $a$ (strong repulsion  for $a_{\mathrm{dd}}/a \ll 1$), as we will see in the following. In this paper, {the } scattering length $a$ (and also $a_{\mathrm{dd}}$) {is} positive.} A  dipolar quasi-2D soliton with an appropriate value of  $a_{\mathrm{dd}}/a$ can be prepared in a laboratory by 
manipulating the value of the scattering length  $a$ by the Feshbach resonance technique \cite{feshbach}. The
use of the quasi-2D GP equation (\ref{eq.GP2d}) will facilitate 
the study of the formation of a strongly dipolar soliton and the collision dynamics of two such solitons.
The quasi-2D dipolar soliton can be studied by a numerical solution of the full 3D 
GP equation (\ref{eq.GP3dHO}). However, it is computationally more economic to solve the quasi-2D equation 
(\ref{eq.GP2d}).

\section{Result: Stationary soliton}

\label{sec3}

We solve the full 3D GP equation (\ref{eq.GP3dHO}) as well as the quasi-2D GP equation (\ref{eq.GP2d}) numerically 
by the Crank-Nicolson discretization algorithm employing the split-time-step propagation method \cite{180CPC09}. 
The  dipolar interaction is treated by a convolution to the Fourier momentum space  \cite{goral} using the available
C and FORTRAN programs \cite{dip}. 
The stationary bound solitons were obtained by  imaginary-time propagation whereas the real-time propagation was applied to study the dynamics {of collision between two quasi-2D solitons. Nevertheless, the imaginary-time method has its limitations. This method may fail when the solution is unstable and the convergence is slow when the solution is close to unstable.  There are the fixed-point schemes, like the variants of Newton-Raphson methods, and improved imaginary-time methods appropriate in these cases.  
{ Also, there are variants to imaginary-time methods,
that are significantly better than
the simple imaginary time methods \cite{yang}. }
These may not be of concern in the present study, as the solitons we will study are stable ground states and the dynamics of collision, often leading to unstable states,  is studied by the real-time method. }

In this paper we will present the results in dimensionless units and take the time scale  $t_0$ to be 1 ms corresponding to the harmonic trap frequency { $\omega_y= t_0^{-1} = 2\pi\times  159.16 $ Hz. The dimensionless results can be converted to actual physical results using the time scale $t_0$ (= 1 ms) and length scale $l_0$. The values of the length scale $l_0$, for this fixed time scale, are different for the commonly used dipolar atoms in BEC experiments, as the masses are different;    
for the dipolar atoms $^{52}$Cr, $^{164}$Dy, and $^{168}$Er the actual values of the length scale  are $l_0\equiv \sqrt{\hbar/m\omega_y}=  1.10 $  $\mu$m,    0.622 $\mu$m, and 0.615 $\mu$m, respectively.
  Hence the present dimensionless results can be transformed into actual physical units using these time and length 
scales for the common dipolar atoms used in BEC  experiments. }

\begin{figure}[t!]
\begin{center}
 \includegraphics[width=.8\linewidth]{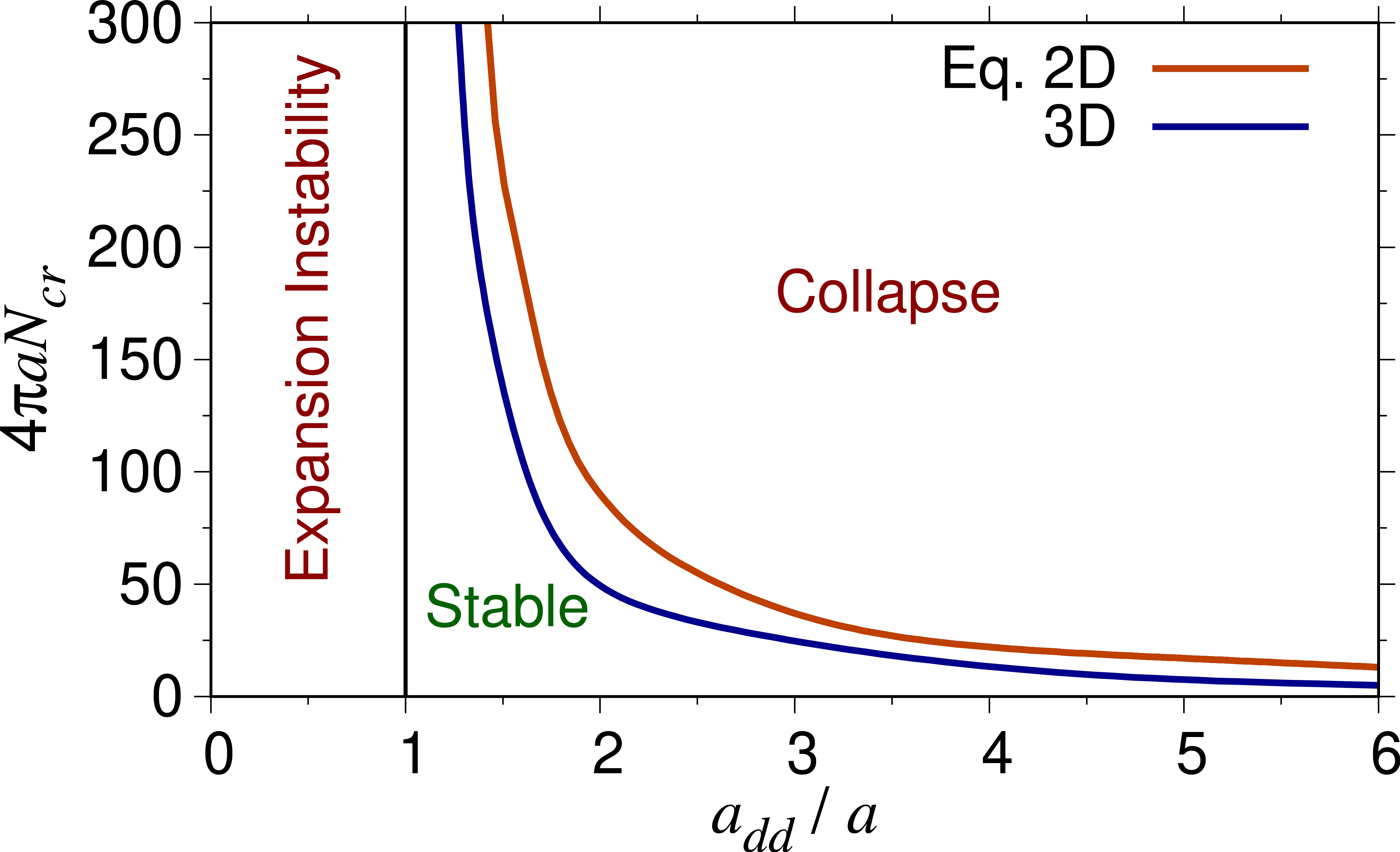}
\caption{ 
Stability phase plot of an anisotropic soliton showing the critical value of interaction { $g_\texttt{cr}\equiv 4\pi a N_{cr}$}
versus $  a_{\mathrm {dd}}/a$
obtained from a numerical calculation using Eqs. (\ref{eq.GP3dHO}) and (\ref{eq.GP2d}) for the 3D and quasi-2D models,
 respectively.  The plotted quantities  in all figures of this paper  are in dimensionless units.
}
\label{fig1} 
\end{center}
\end{figure}


As it is commonly known,
a BEC with dipolar interaction always becomes unstable and collapses when the number of particles is sufficiently large
and greater than a critical number $N_{\mathrm{cr}}$ ($N > N_\texttt{cr}$), independent 
of the trapping geometry \cite{dicol}. 
This is also true for the quasi-2D soliton described by Eq. (\ref{eq.GP3dHO}) in 3D and 
Eq. (\ref{eq.GP2d}) in 2D.
A stability plot for the quasi-2D dipolar soliton 
is  illustrated in Fig. \ref{fig1} from numerical solutions of  the 3D equation (\ref{eq.GP3dHO}) and the quasi-2D equation (\ref{eq.GP2d}) by imaginary-time propagation. {The two results are in agreement with each other,  specially for larger values of $a_{\mathrm{dd}}/a$, where the soliton acquires a quasi-2D shape and validates the use of the quasi-2D equation (\ref{eq.GP2d}).}
The stability phase diagram in Fig. \ref{fig1} is consistent with 
previous results 
 obtained by Santos \textit{et al.} 
\cite{102PRL09}.
In the stable region, a balance between repulsive and attractive interactions allows the formation of stable solitons. 
However, in the expansion region, attractive dipolar interaction is not enough to compensate for the atomic contact repulsion; 
consequently, the soliton cannot be bound and will escape to infinity.
In the collapse region the dipolar attraction is much larger than the contact repulsion and the soliton collapses.  
 
\begin{figure}[t!]
\begin{center}
\includegraphics[width=.7\linewidth]{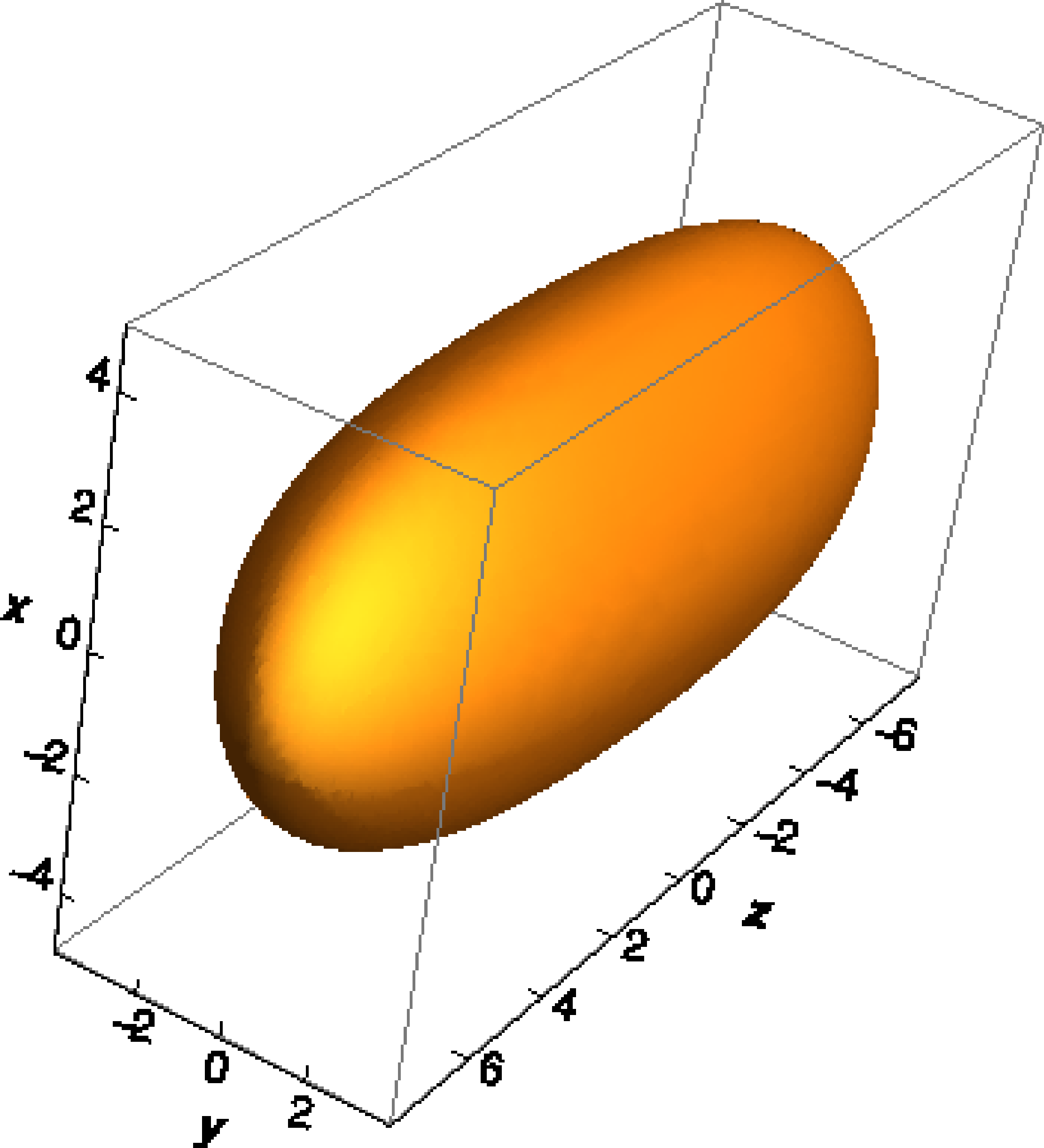}
\caption{The 3D contour plot of  a  anisotropic quasi-2D dipolar soliton with $g=16$ and  $a_{\mathrm{dd}}/a=3.35$ in dimensionless units, obtained from a solution of the 3D GP equation (\ref{eq.GP3dHO}). 
 The density on the contour is 0.001. 
}
\label{fig2} 
\end{center}
\end{figure}

\begin{figure}[t!]
\begin{center}
 \includegraphics[width=\linewidth]{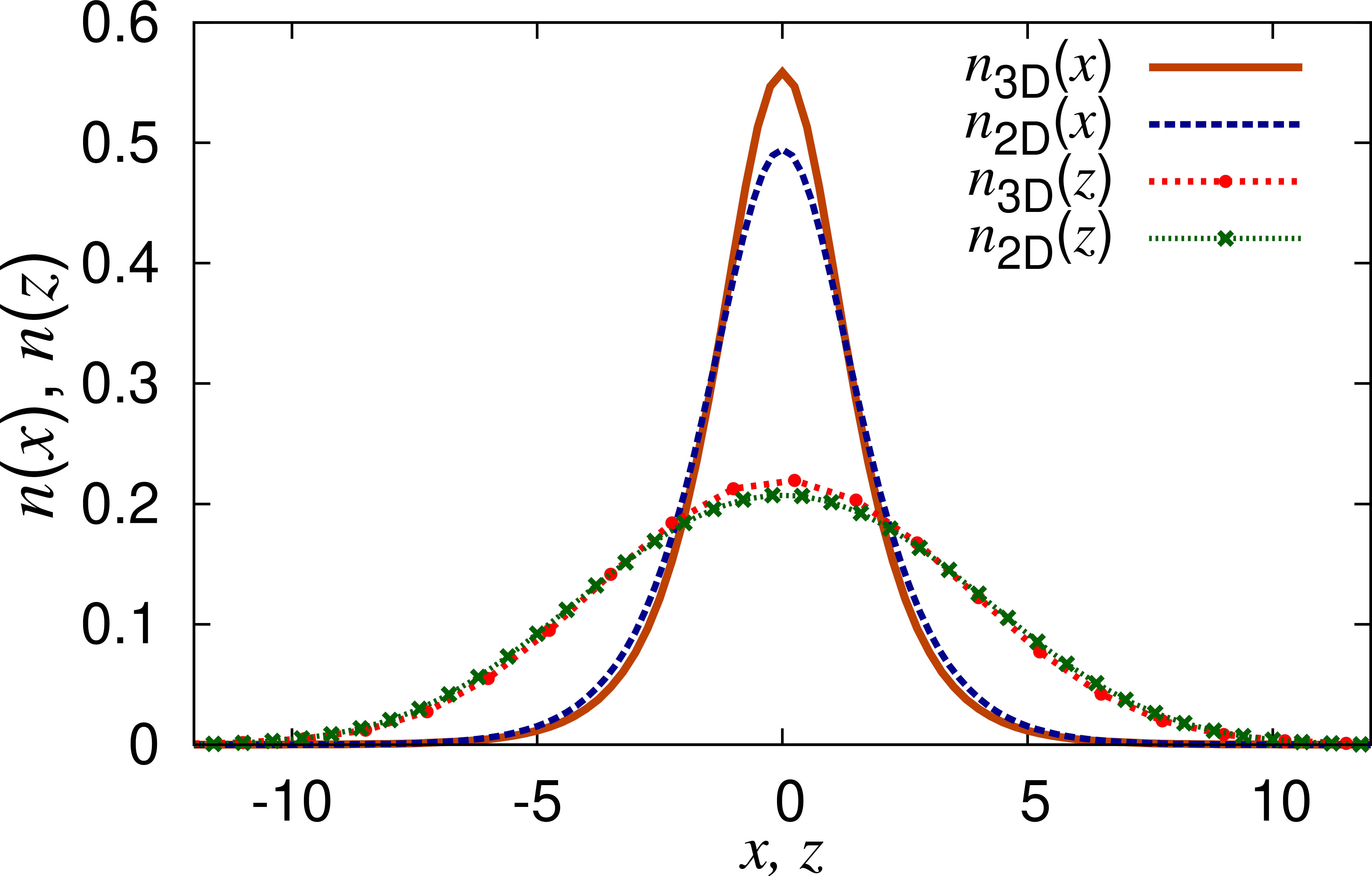}
\caption{Reduced densities $n_{3D}(x), n_{3D}(z),n_{2D}(x),$ and $n_{2D}(z)$ versus $x$ and $z$ in dimensionless units 
for the quasi-2D soliton of Fig. \ref{fig2}
obtained 
from the 3D and quasi-2D models (\ref{eq.GP3dHO}) and (\ref{eq.GP2d}).  
}
\label{fig3} 
\end{center}
\end{figure}

In
Fig. \ref{fig2} we illustrate the shape of a stable anisotropic dipolar quasi-2D soliton showing the 3D contour plot of density 
obtained from  a numerical solution of the 3D GP equation (\ref{eq.GP3dHO}) by imaginary time propagation \cite{180CPC09}
for $g=16$ 
and
\mbox{$a_{\mathrm{dd}}/{a}=3.35$}. Because of the harmonic trap in the $y$ direction, the quasi-2D 
soliton is extended in the $x-z$ plane with a small width in the $y$ direction. 
{ The actual values of the ratio \mbox{$a_{\mathrm{dd}}/{a}=3.35$} for the commonly used dipolar atoms is smaller than 3.35. For example, 
for $^{164}$Dy,   $a_{\mathrm{dd}} =131a_0,$  $a=92a_0$ \cite{dy}, $a_{\mathrm{dd}}/a \sim 1.42$,  for $^{168}$Er $a_{\mathrm{dd}} =65a_0$,  $a=137a_0$, $a_{\mathrm{dd}}/a \sim 1.47$ \cite{er}, 
for $^{52}$Cr  $a_{\mathrm{dd}} =16a_0$ \cite{dipbec}   $a\sim 100a_0$ \cite{cr}  $a_{\mathrm{dd}}/a \sim 0.16$ \cite{cr52}, where $a_0$ is the Bohr radius.   A smaller value of $a$, and consequently,  a larger $a_{\mathrm{dd}}/a $, 
gives a quasi-2D shape of the soliton  with small spatial extension along the perpendicular $y$ direction.  In a laboratory a smaller value of $a$ can be achieved by the Feshbach resonance technique \cite{feshbach}. Such a modification of the atomic scattering length by a Feshbach resonance  was necessary in all BEC  experiments on solitons \cite{417N02,solrb,hulet}. 
}

For an anisotropic quasi-2D soliton confined in the $x-z$ plane, it is appropriate to compare the reduced  densities along $x$ and $z$ directions 
obtained from the 3D equation (\ref{eq.GP3dHO}) and the quasi-2D equation (\ref{eq.GP2d})  by integrating the respective densities over the orthogonal coordinates. The reduced  densities are calculated from the 2D densities  as
\begin{align}\label{n2d}
n_{2D}(x)&=\int |\phi({\boldsymbol \rho})|^2 dz, \quad 
n_{2D}(z)&=\int |\phi({\boldsymbol \rho})|^2 dx,
\end{align}
and from the 3D densities as
 \begin{align}
n_{3D}(x)&=\int |\psi({\bf r})|^2 dz dy,\quad 
n_{3D}(z)&=\int |\psi({\bf r})|^2 dxdy.
\end{align}
In Fig. \ref{fig3} we compare these axial densities of the quasi-2D soliton of Fig. \ref{fig2}
obtained from the 3D and 2D equations. 
Based on the densities displayed in Fig. \ref{fig3} and the stability plots of Fig. \ref{fig1}, we find that the
2D reduction Eq. (\ref{eq.GP2d}) provides a faithful account of the
actual state of affairs, 
when compared to the full 3D Eq. (\ref{eq.GP3dHO}).

\section{Result: Collision Dynamics}

\label{sec4}

Starting with the initial anisotropic soliton wave function obtained by imaginary-time propagation we
can obtain subsequent dynamics generated by real-time propagation of 
Eq. (\ref{eq.GP2d}). { The GP equation is Galilean invariant, hence a moving soliton, necessary for studying the collision dynamics, can be generated in a routine fashion in numerical simulation. A moving soliton traveling with a velocity $\bf v$  can be trivially obtained in real-time simulation using a converged stationary solution  of imaginary-time propagation multiplied by the phase factor $\exp(i \bf v \cdot  \boldsymbol \rho)$ 
as the initial function, provided an infinitely small space and time steps are used in numerical simulation.
}

\begin{figure}[t]
\begin{center}
 \includegraphics[width=\linewidth]{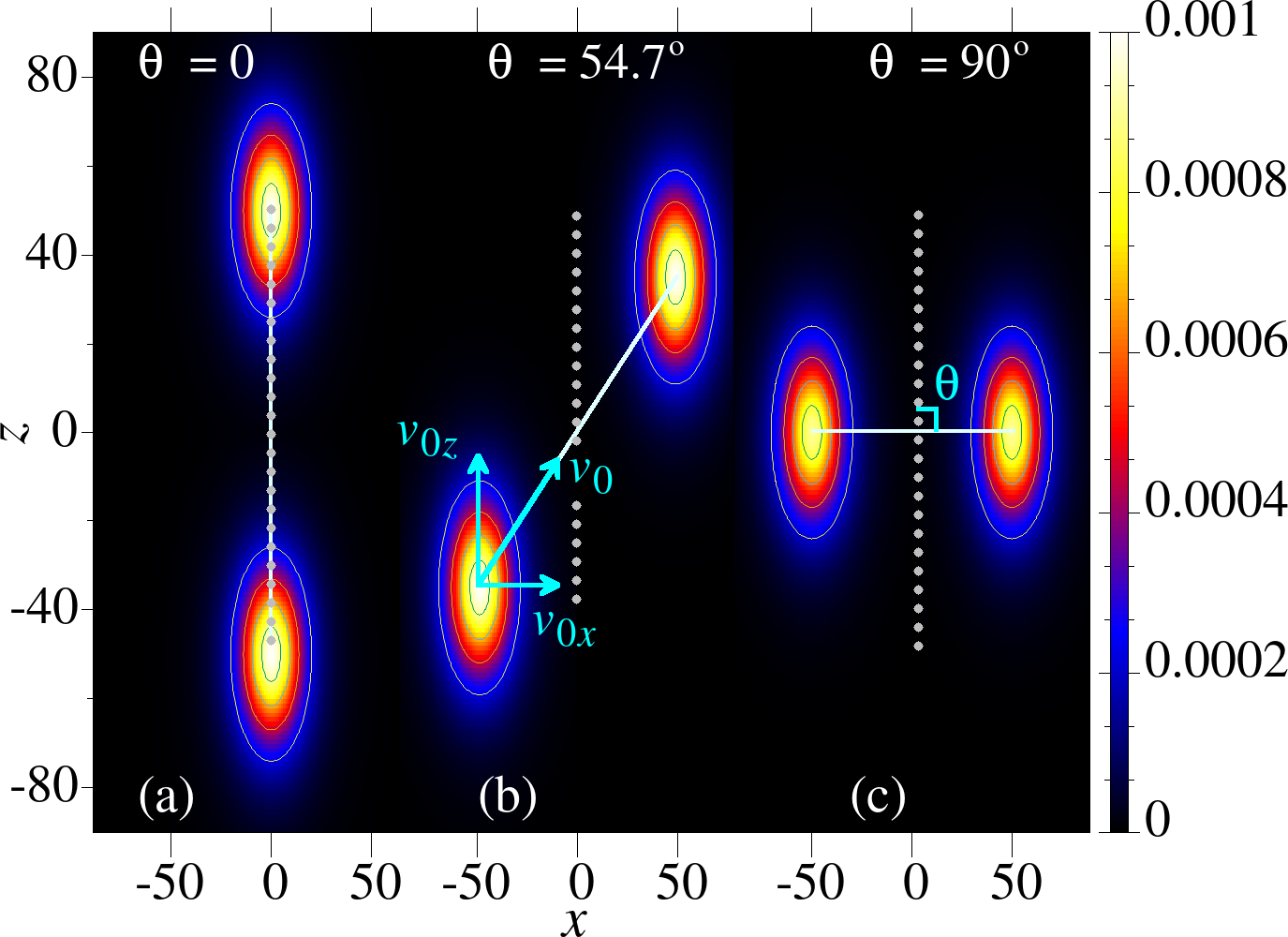}
\caption{ (a) A generic initial configuration of the two colliding  anisotropic quasi-2D dipolar solitons  with $g=8$ and $a_{\mathrm{dd}}/a=3.35$ through a
contour plot of density  $|\phi(x,z)|^2$.  The solitons are moving towards each other with velocity $\pm {\bf v}_0$ with  the vector ${\bf v}_0$ making an angle $\theta$ with the $z$ axis { (gray dot line)}.  
The { white full} lines joining the two solitons  represent the collision directions for 
(a) $\theta=0$,  
({b}) $54.7\degree$ and 
({c}) $90\degree$.
}
\label{fig4} 
\end{center}
\end{figure}

Two such solitons, prepared by imaginary-time propagation, are placed in different 
spatial orientation  at positions $\pm \boldsymbol \rho $ and time $t=0$ in the $x-z$ plane modifying the initial angle ($\theta$) between 
the collision direction and the axis of polarization $z$. { The respective imaginary-time wave functions are then multiplied by appropriate  phase factors  $\exp(\pm i \bf v \cdot  \boldsymbol \rho)$
and then used as the initial states of real-time propagation so as to simulate the collision dynamics of two solitons at $\pm \boldsymbol \rho$  colliding frontally at the center $\boldsymbol \rho=0$ with velocity $\pm \bf v$.} 

A generic initial configuration of frontal collision of  two dipolar  solitons moving towards each other with 
velocity $\pm {\bf v}_0$ of components $v_{0x}$ and $v_{0z}$ is shown through  contour density plots in   Fig. \ref{fig4}.
Three different initial configurations of the two colliding solitons  for $\theta =0, 54.7 \degree $ and $90 \degree$ are 
shown in { Fig. \ref{fig4} (a)-(c)} through a contour plot of the respective densities.
Of these, the orientation $\theta =54.7 \degree $ 
is interesting as strictly along this direction the interaction  between two dimensionless point  dipoles is zero \cite{dipbec}. { The effective interaction between two solitons is the folding integral over the  interatomic   interactions between the atoms in the two solitons.
Consequently,  in the asymptotic region for angle $\theta=54.7$\degree, when the two solitons are well separated, the effective  dipolar  interaction will be small, as the interatomic interactions around this angle will be small.  For angle $\theta =0$, the effective dipolar  interaction between the two solitons  will be repulsive as two point dipoles placed in side-by-side position {repel} \cite{dipbec}. For angle $\theta =$ 90\degree, the effective dipolar  interaction is attractive as two dipoles placed in head-to-tail position attract each other. Hence these three angles correspond to distinct dipolar 
asymptotic interactions between the solitons.  For all these angles the isotropic nondipolar contact interaction in the asymptotic configuration is zero.}
In this way we study, in real-time simulation, the collision of two 
identical { in-phase}
solitons placed as in Fig. \ref{fig4}.  {In quasi-elastic collision the emerging solitons are also in phase.} 
We can use the variation in the initial conditions by varying the parameters 
 $\theta$ and ${\bf v}_0$ 
to explore the general dynamics involved in the collision process between two anisotropic dipolar quasi-2D solitons.

 \begin{figure}[!t]
\begin{center}

 \includegraphics[width=.95\linewidth]{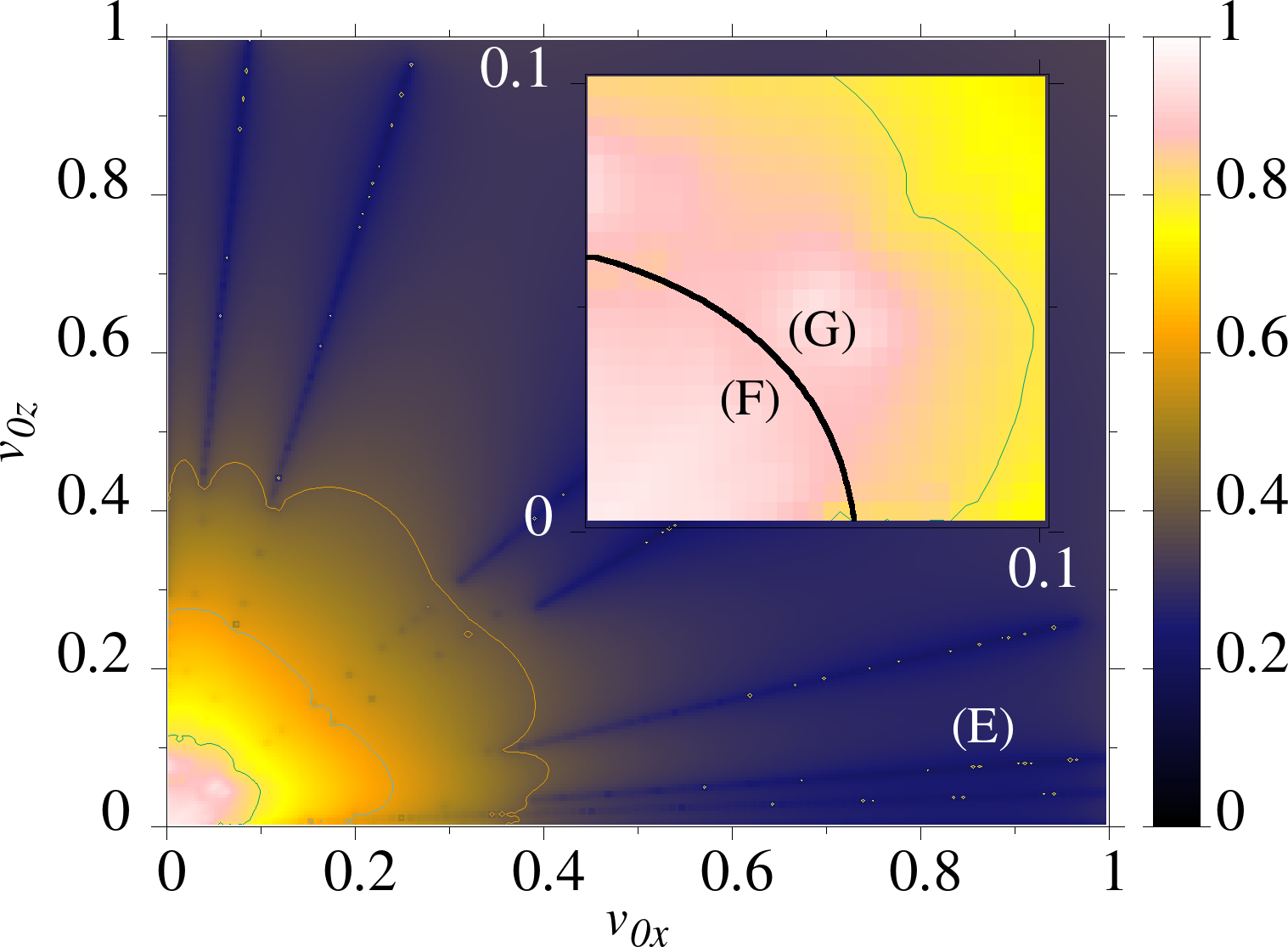}
\caption{ 
  Variation of the soliton velocity ${\boldsymbol \Delta}{\bf  v}/v_0$ as a function of initial velocity 
$v_0$ for different angles $0\leq \theta \leq 90\degree$. The collision is quasi elastic for $\Delta v/v_0 \simeq 0$ in region (E). 
Inset: Region of deep inelastic collision  with the  formation of a coalesced soliton for $\Delta v/v_0 \simeq 1$:    unexpected change in the 
direction of propagation (full black line), 
 scissors-mode oscillation in region (F),  quadrupole-mode excitation in region (G). {The colorbar represents the value of $\Delta v/v_0$ in the plot for a particular color. }
}
\label{fig5} 
\end{center}
\end{figure}

 The complex dynamic processes involved in the collision between two anisotropic solitons 
in a dipolar BEC, which depend on the angle $\theta$, is controlled by the velocity vector ${\bf v}_0$ or by  its two  components 
$ v_{0x}$  and $ v_{0z}$. The similar collision dynamics between two isotropic nondipolar collision is controlled  only by the magnitude of the velocity vector $v_0=\sqrt{v_{0x}^2+v_{0z}^2}$ and not by the components $ v_{0x}$  and $ v_{0z}$,  being independent of the colliding angle $\theta$. To identify the domains of elastic and inelastic collisions, 
it is convenient to define the relative variation of soliton velocity vector ${\boldsymbol \Delta} {\bf v}/v_0 = [{\bf v}_0(t=0) -{\bf v}(t\rightarrow \infty)]/v_0$ and study its variation in the  collision process  
for different angles $\theta $. 
 In Fig. \ref{fig5} we elaborate a   contour plot of the variation of the soliton velocity ${\boldsymbol \Delta} {\bf v}/v_0 $
  for collision of two identical solitons with $g=8$ and $a_{\mathrm{dd}}/a=3.35$ each 
as a function of the initial velocities $v_{0x}$  and $v_{0z}$ {and of}
 different angles $0\leq \theta \leq 90\degree$. { The angle $\theta$ 
in this plot is inherent in the sizes of $v_{0x}$ and $v_{0z}$. For example,
in this plot $v_{0x}=0$ corresponds to $\theta = 0$  and  $v_{0z}=0$  to $\theta =90$\degree.}
Obviously, ${ \Delta} {v}/v_0=1$
{denotes} fully inelastic collision { with a total loss of the kinetic energy to inelastic excitation}
and ${  \Delta} { v}/v_0=0$ {denotes} elastic collision {with the conservation of kinetic energy}.
As expected, during soliton-soliton collision processes, the total energy of the system is conserved \cite{76PRA07}. 
For large velocities in the region (E) of Fig. \ref{fig5}
the collision may be considered as quasi elastic and  $\Delta v/v_0 \simeq 0$ (black color in colorbar).
However, for small velocities {near the origin  with $\Delta v/v_0 \simeq 1$ (pink color in colorbar)}, 
the collision becomes inelastic. 
 {These different possibilities of inelastic excitation are further displayed in details in the inset of Fig. \ref{fig5} for small $v_0$.}

\begin{figure}[t]
\begin{center}
 \includegraphics[width=\linewidth]{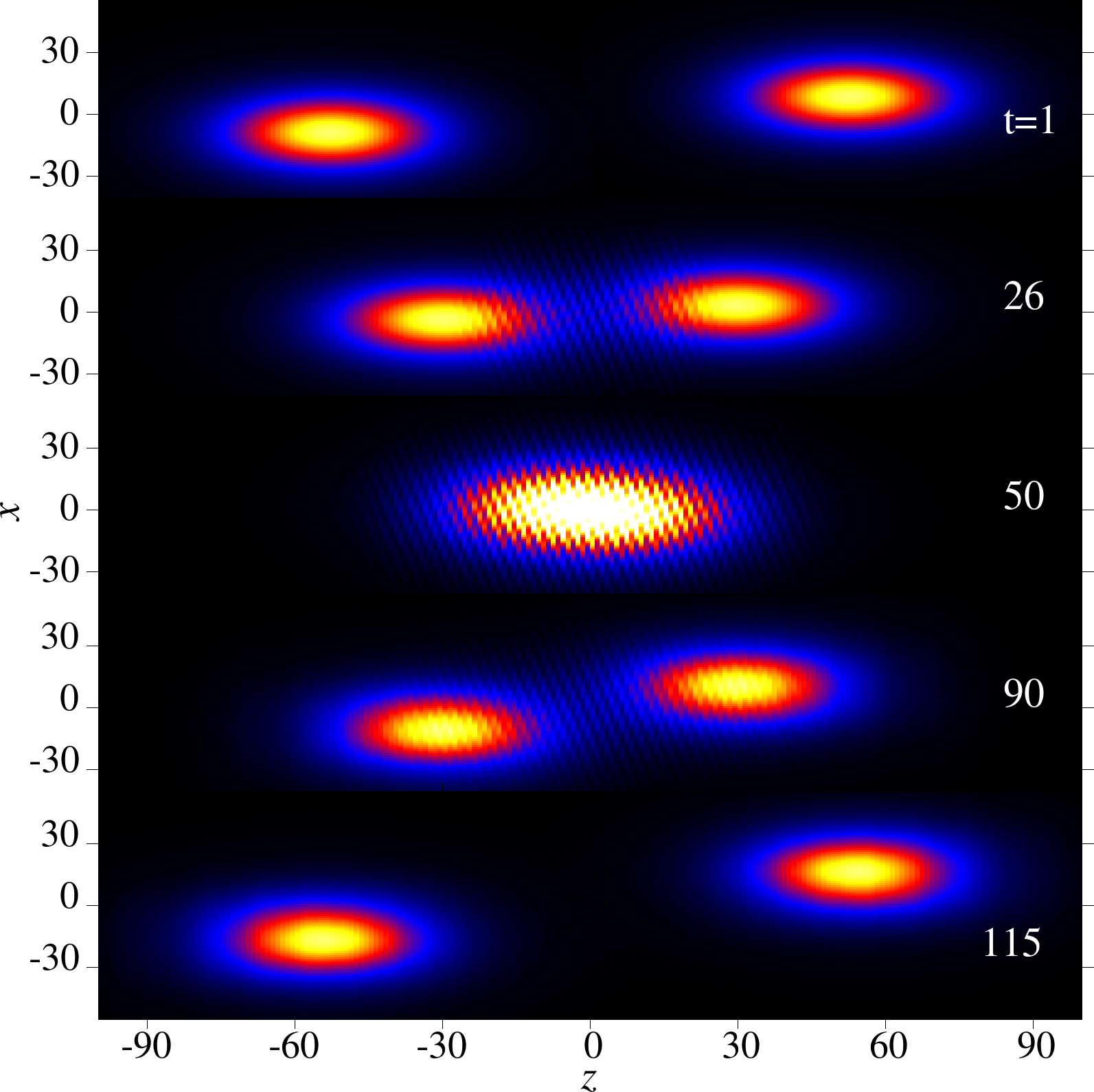}
\caption{Contour plot of density $|\phi(x,z)|^2$ for the
collision dynamics of two solitons  with $g=8$ and $a_{\mathrm{dd}}/a=3.35$ each 
 initially placed at $z=\pm51$, {with 
$\theta=10\degree$}, and  moving with velocity $v_0 =\pm 1$  
before ($t$ = 1 and 26 ), during ($t$ = 50) and after ($t$ = 90 and 115) collision
in a quasi-elastic process (same color bar as in Fig. \ref{fig4}). The solitons come towards each other, interact at $z=0$, and then separate and come out practically unchanged.
}
\label{fig6} 
\end{center}
\end{figure}

Although at large velocities the collision between two solitons is quasi elastic {with the conservation of  total kinetic energy,
as the velocity of the colliding solitons is reduced, the collision between two solitons gradually {becomes} inelastic with a deformation in the shape of two emerging solitons. In some of these inelastic collision processes at very small velocities we found evidence of coupling with special internal soliton modes, which will be called a deep inelastic process,
such as scissors mode, quadrupole mode and even an 
interesting effect which produces an unexpected change in the direction of propagation {of the emerging solitons}.  With a reduction in velocity, 
the incoming  solitons may form a coalesced soliton in an excited  state, which may break up into more than two pieces.  In this region we identified    an excitation to a    dipole or quadrupole mode, viz. region (G) in Fig. \ref{fig5}. For even smaller velocities the excited coalesced soliton does not have enough energy to break up into multiple pieces, but eventually decays into two solitons that move apart in opposite directions distinct from the direction of motion of the incident soliton, viz. black line  in the inset of  Fig. \ref{fig5}.}
For sufficiently small velocities, the  coalesced soliton does not have enough energy to break up and remain in an excited  oscillating state, often  { undergoing oscillation in scissors mode or in monopolar breathing mode, viz. region (F) in Fig. \ref{fig5}. For the demarcation between regions (F) and (G), we {performed} simulation for collision for ten different angles between $\theta=0$ and 90\degree.}

\subsection{Head-on quasi-elastic collision}
\label{4.1}

\begin{figure}[!t]
\begin{center}
\includegraphics[width=\linewidth]{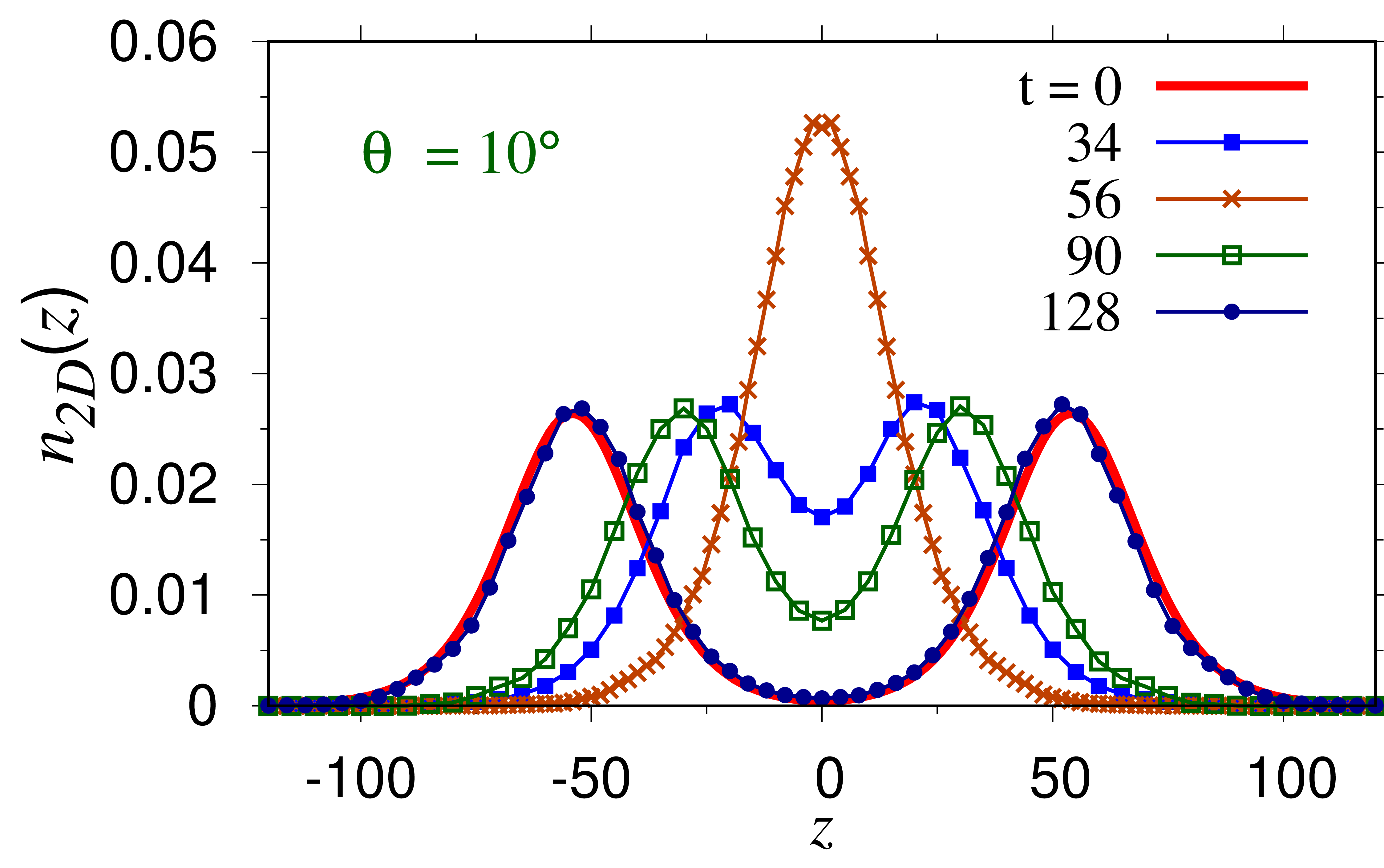}
\caption{ 
Reduced density $n(z)=\int |\phi(x,z)|^2 dx$ in a quasi-elastic process of Fig. \ref{fig6}
before ($t$ = 0 and 34), during ($t$ = 56) and after ($t$ = 90 and 112) collision.
Very similar results 
for initial ($t=0$) and final ($t=112$) densities demonstrate the quasi-elastic nature 
of collision.  
}
\label{fig7} 
\end{center}
\end{figure}

Only the collision between two integrable 1D solitons is truly elastic. However, 
for  sufficiently large initial velocities, the collision between two quasi-2D dipolar 
solitons  is  found to be  quasi-elastic with no visible change in shape and {also in velocity after the collision, preserving the kinetic energy.}  
In this case,
the relative distance between the two centers of mass increases with time after the collision.
We consider the head-on collision between two solitons, initially placed at $z=\pm 51$ at an angular orientation $\theta =10 \degree$ and attributed a velocity of $v_0=\pm 1$  so as to collide frontally. 
In  Fig. \ref{fig6})  we illustrate the quasi-elastic nature of this  collision using 
the contour plot of 2D density $|\phi(x,z)|^2$,   through snapshots of density  at different times 
 before ($t= 1$, 26), 
during ($t=$ 50) and after ($t=$ 90, and 115) collision.
{ The quasi-elastic nature of collision becomes more explicit from a plot of the reduced density $n_{2D}(z)$ during collision.}
 The  reduced densities  $n_{2D}(z)$, viz. Eq. (\ref{n2d}), of the colliding solitons  
  at different times before ($t= 0$, 34), 
during ($t=56$) and after ($t=90$,  128) collision are displayed in Fig. \ref{fig7}. {  We find 
in Fig. \ref{fig7}
that the reduced densities at time $t=0$ and $t=128$, before and after collision, respectively, are practically the same. 
 The quasi-symmetric profile of   $n_{2D}(z)$ during collision demonstrates that }
 the solitons come out practically unchanged after 
collision and the quasi-elastic nature of collision is confirmed from a comparison of densities at $t=0$ and 
$t=128 $ in Fig. \ref{fig7}.
A video clip  of the collision dynamics of Fig. \ref{fig6} 
is prepared and contained in supplementary 
file head-on-quasi-elastic-collision.mp4 and also the video available at 
 {\color{blue} https://youtu.be/BnZLlVsIGe0} .

\begin{figure}[t]
\begin{center}
 \includegraphics[width=\linewidth]{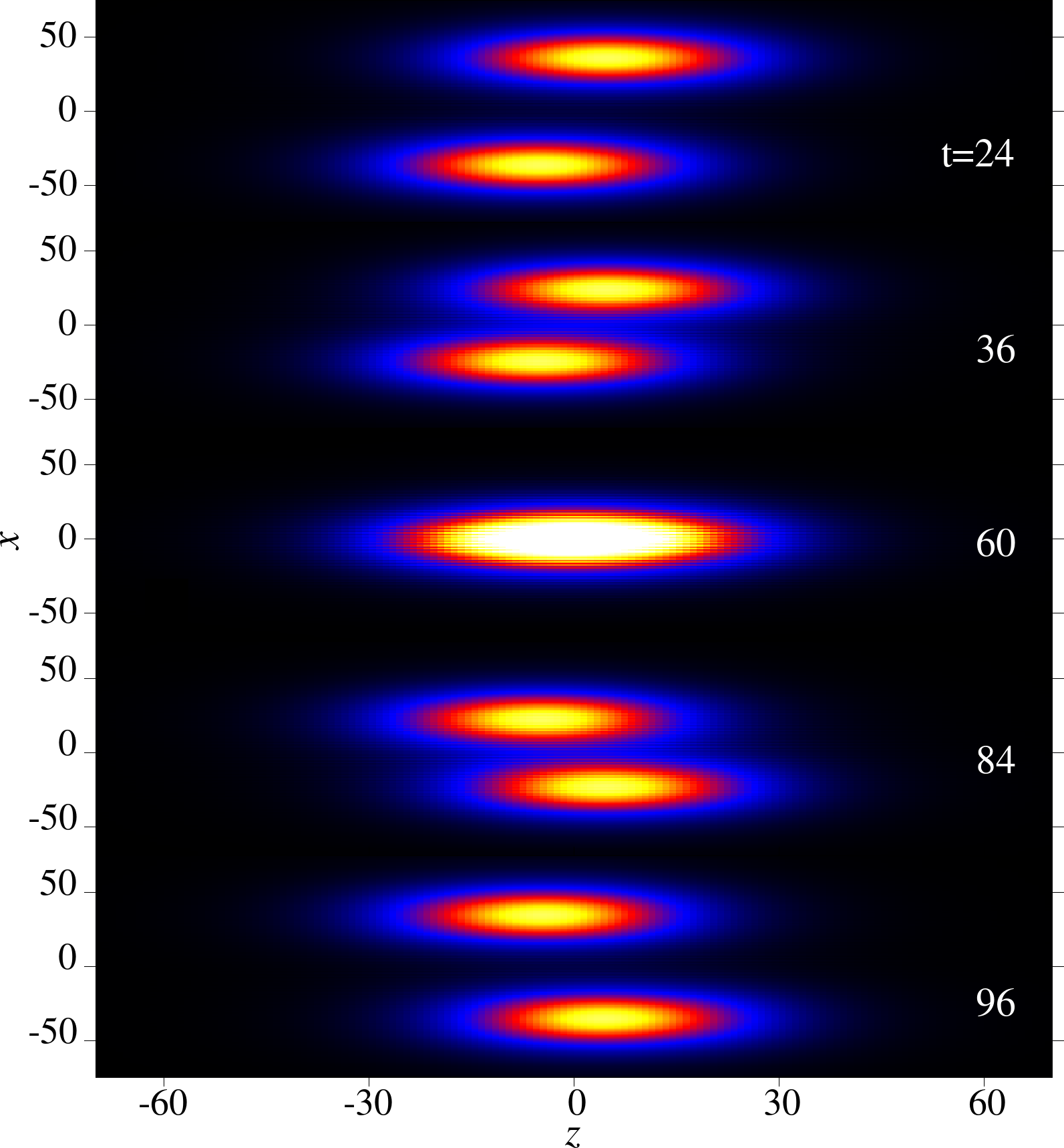}

\caption{Contour plot of density $|\phi(x,z)|^2$ for the
collision dynamics of two solitons  with $g=8$ and $a_{\mathrm{dd}}/a=3.35$ each 
 initially placed at {$x=\pm 60$ and $z=\pm 5$,} {with 
$\theta=90\degree,$} and  moving with velocity {$v_0 =\pm 1$}  with an impact parameter of 10
before ($t$ = 24 and 36), during ($t$ = 60) and after ($t$ = 84 and 96) collision
in a quasi-elastic process (same color bar as in Fig. \ref{fig4}). The solitons come towards each other, interact at $x=0$, and then separate and come out practically unchanged maintaining the initial directions of motion.
}
\label{fig8} 
\end{center}
\end{figure}

{

\subsection{Quasi-elastic collision with an impact parameter}
\label{4.2}

We also studied quasi-elastic collision with a non-zero impact parameter and moving along the $x$ direction.  For this we considered two solitons   placed at {$x= \pm 60$ and $z= \pm 5$} at $t=0$ and attributed the velocity {$v_0 =\pm  1$ }
along the $x$ direction with $\theta =90$\degree so as to collide at $x=0$ with an impact parameter of 10.
In this case  the soliton continues to  move along the $x$ direction after collision without any deformation in shape or notable change of velocity or direction of motion along $x$ direction.   {Such a collision could introduce} a rotation in the solitons after collision or a  change in the direction of motion.   No such rotation is found in this
case or change in the direction of motion, viz. Fig. \ref{fig8}. We performed numerical simulation for different values of the impact parameter but the outcome remains unchanged.   
A video clip  of the collision dynamics of Fig. \ref{fig8} is contained in supplementary file
quasi-elastic-with-impact-parameter10.mp4 and also video  available at 
{\color{blue} https://www.youtube.com/watch?v=5AKpui0tlYY} . }

\subsection{Deep inelastic collision}

\label{4.3}

{As the velocity is reduced the collision becomes inelastic and the emerging solitons after collision are deformed in shape but maintain the initial directions of motion. Upon further reduction in velocity, the two colliding solitons form a coalesced soliton in a highly excited state performing oscillation, which eventually breaks up, often into more than two pieces. In this case the coalesced soliton 
does not ``remember" the incident directions of motion and the emerging solitons come out in directions independent  of the initial directions.   The collision in this case (with velocities corresponding to the inset of Fig. \ref{fig5}) will be called deep inelastic collision. In the following we  consider three distinct phenomena  encountered in deep inelastic collision. The process of deep inelastic collision 
is quite similar to compound nuclear reaction \cite{nbook}.  In compound nuclear reaction, two low-energy colliding nuclei form a compound nucleus in an excited state, which do not remember the details of the incident nuclei and eventually breaks up into different pieces which come out in directions independent of those of the directions of motion of the incident nuclei.   }

\subsubsection{Scissors-mode oscillation {and soliton fusion}}

\begin{figure}[!t]
\begin{center}
\includegraphics[width=\linewidth]{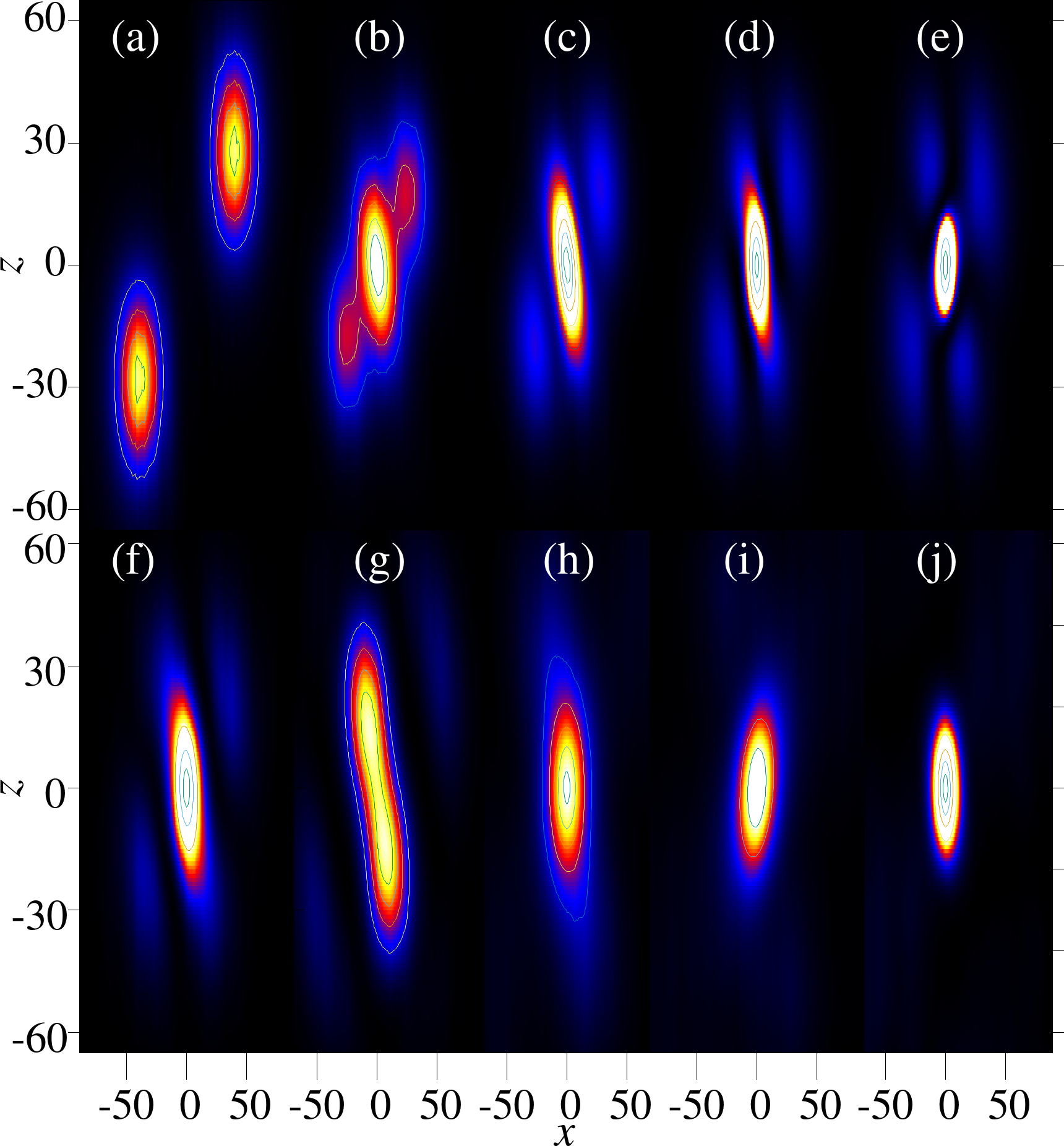}
\caption{ 
Contour plot of densities $|\phi(x,z)|^2$ 
in a collision with an initial angle of $\theta=54.7\degree$ {between the direction of motion and the direction of polarization $z$}
and velocity $v_0=0.04$ at different times   before (a)  $t=40$, during  (b) $t=680$,  and  after  (c)  $t=800$, (d) $t=920$, (e) $t=1040$, (f)  $t=1160$, (g)  $t=1400$, (h) $t=1920$, (i)  $t=2160$ and (j) 2640    collision.
The initially independent solitons become bound into a new excited coalesced state  in
(c)-(j) oscillating around the direction of polarization  and 
eventually   fully oriented along  the $z$ axis  (color bar as in Fig. \ref{fig4}). 
}
\label{fig9} 
\end{center}
\end{figure}

\label{4.2.1}

We next study  the formation of a coalesced soliton executing scissors-mode oscillation, viz. region (F) in Fig. \ref{fig5}, 
in deep inelastic collision  of  two anisotropic quasi-2D dipolar solitons, each 
with $g=8$ and $a_{\mathrm{dd}}/a =3.35$, placed at $z=\pm 29$, with an 
initial angle $\theta=54.7\degree$ 
between the collision direction and the axis of polarization $z$
and with an initial velocity of $v_0=0.04$  at $t=0$. In this angular orientation the dipolar interaction between two 
point dipoles is zero and that between the two solitons is supposed to be small when separated by a distance large compared to their size. 
We plot in Fig. \ref{fig9} the contour density $|\phi(x,z)|^2$ at different times
before (a)  $t=40$, during  (b) $t=680$,  and  after  (c)  $t=800$, (d) $t=920$, (e) $t=1040$, (f)  $t=1160$, (g)  $t=1400$, (h) $t=1920$, (i)  $t=2160$ and (j) 2640    collision.   {The formation of the  coalesced soliton is complete in plot (c) at time $t=800$. 
The initial orientation of the prolate solitons in (a) is along the $z$ axis. The slowly colliding solitons  merge into a single coalesced soliton in (c), whose orientation is different from the original (vertical) one.
We find in (b), (c) and subsequent plots two small separated pieces  resembling the density of an angular momentum $L=1$ state usually called a dipole excitation, which eventually subsides.

In the deep inelastic collision of two anisotropic quasi-2D dipolar solitons we find the signature of 
scissors-mode oscillation.
The orientation of the prolate soliton after formation in Fig. \ref{fig9} also performs a few angular 
 scissors-mode 
oscillations,  and eventually {forms a fused soliton, which}  relaxes   along the polarization  $z$ axis in (j) due to the dipolar interaction. 
{ During this process the kinetic energy of oscillation is changed to internal excitation energy of the fused soliton.} 
We illustrate this scissors mode in the dynamic collision using 
the Figs. \ref{fig9} (c)-(i) to show the oscillation of density cloud  around the $z$ axis, 
and after some time, Fig. \ref{fig9} (j), the system is fully oriented along the direction of polarization.
In  plots (c) to (e) the soliton turns to right,  and then turns to left  in plots (e) to (g), then to right in plots (g) to (i) and then to  left to relax along the polarization $z$ direction in plot (j).

\begin{figure}[!t]
\begin{center}
\includegraphics[width=\linewidth]{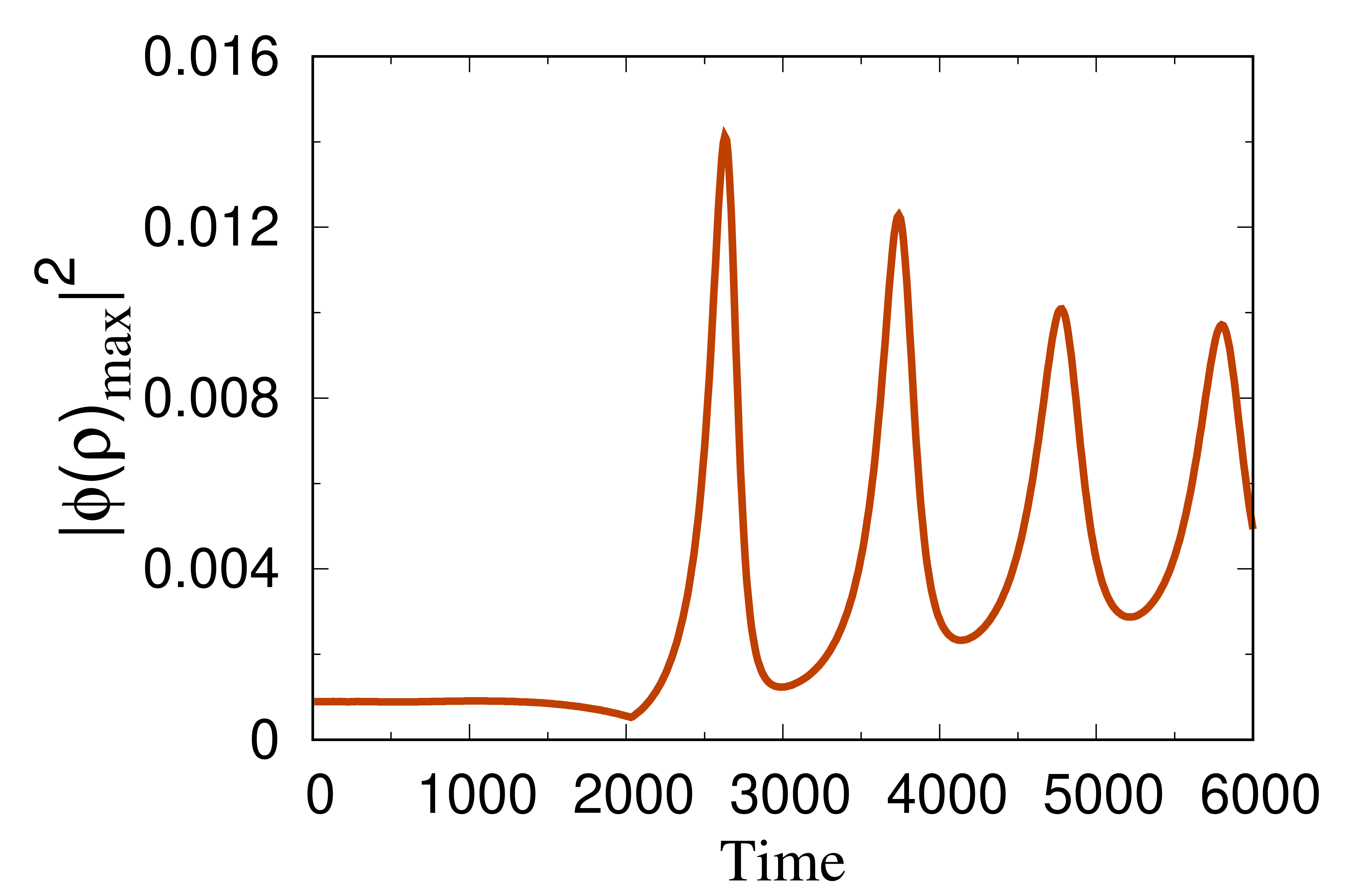}
\caption{ 
The maximum value of density $|\phi(x,z)_{\text max}|^2$ vs time
for the inelastic process in Fig. \ref{fig9}.
The value of density is oscillating in time after collision.  
}
\label{fig10} 
\end{center}
\end{figure}

As a coalesced soliton aligns itself eventually along the polarization direction because of dipolar interaction,  it is not possible to have the final coalesced soliton   aligned in a direction  different from the $z$ axis. 
After formation, the coalesced soliton shrinks in size  and its central density attains a maximum and the soliton continues a monopole breathing oscillation (radial contraction and expansion) for a long time.
To demonstrate this oscillation  explicitly, we plot in  Fig. \ref{fig10} the maximum value of density after  collision versus 
time. {We have continued the  plot in  Fig.  \ref{fig10} for times  well beyond  that shown in Fig. \ref{fig9}.} The maximum value of density is small in the beginning and attains a maximum value during collision between 
time $t=2000$ and $t=3000$, { i.e., after the formation of the coalesced soliton. Then for   $t>3000$} the maximum density keeps on oscillating as the coalesced
soliton executes a breathing oscillation {during a very long period of time with a period of approximately 1000 time units and with a decaying amplitude.}  
The continued oscillation of density  
 demonstrates the robustness of this new coalesced state for large time.} A video clip of the fusion dynamics of Figs. \ref{fig9} and \ref{fig10}, to show the breathing-mode and scissors-mode oscillations, 
is also prepared and contained in  supplementary file scissors-mode.mp4 and  
 also available at  {\color{blue} https://youtu.be/LB8b8JKOmKA} .
%





\subsubsection{Change in direction of propagation}
\label{4.2.2}

\begin{figure}[!t]
\begin{center}
 \includegraphics[width=.9\linewidth]{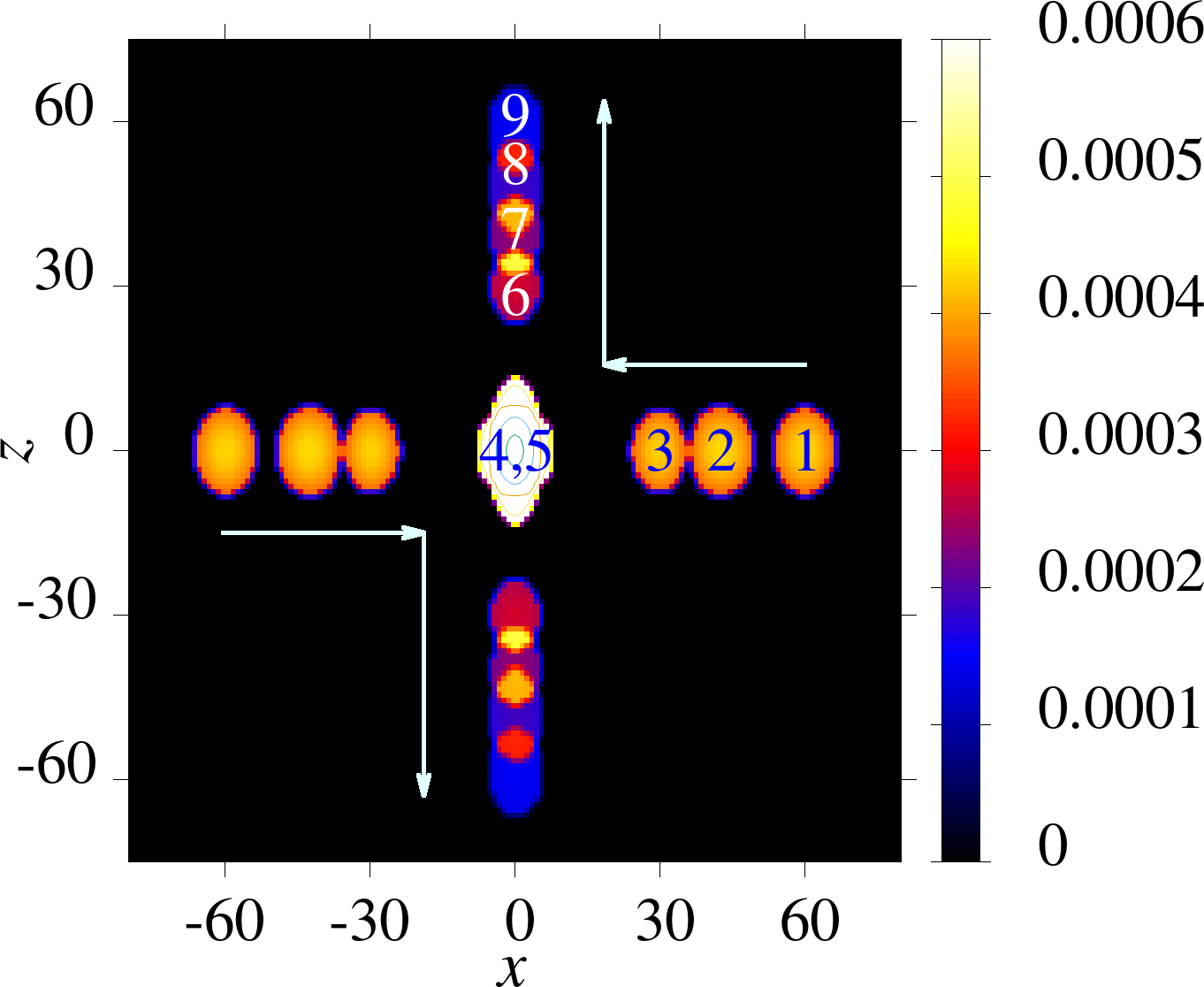}
\caption{ 
Contour plot of density of  two colliding quasi-2D dipolar  solitons with $g=8$ and $a_{\mathrm{dd}}/a=3.35$ each
during inelastic collision 
with initial $\theta=90\degree$ and velocity $v_0=0.06$. 
The numbered snapshots, representing the evolution of density, are  plotted after the fixed interval of time $\Delta t =280$ at times $t=0, 280, 560, 840, 1120, 1400, 1680, 1960, 2240$ and 2520. 
After the collision, the angle between the 
propagation direction and the $z$ axis suddenly changes to the final value of about $\theta_f=0$. 
}
\label{fig11} 
\end{center}
\end{figure}

Interestingly, some of the deep inelastic collision  processes show an unexpected change in the direction of propagation 
for both solitons after the collision: viz. the  full black line in the inset of  Fig.\ref{fig5}.
{This is a classic example of deep inelastic collision, where the final emerging solitons come out in directions completely independent of the directions of the incoming solitons. }
We illustrate in Fig. \ref{fig11}, a collision process for two anisotropic quasi-2D solitons initially placed at $x=\pm70$, 
with initial velocity of about $v_0=0.06$ and an
initial angle of $\theta=90\degree$ between the collision direction and the $z$ axis at $t=0$. 
The contour of density $|\phi(x,z)|^2$ at different times before (1-3), during (4 and 5) and after (6-9) collision in Fig. \ref{fig11}
is showing that the direction of propagation suddenly changes.
{ This happens due to the formation of a coalesced soliton in an highly excited state. Due to an excessive energy, this soliton eventually decays into two solitons which move apart in opposite directions, distinct from the incident direction.}
The arrows in Fig. \ref{fig11} represent the direction of propagation
for both solitons before and after collision showing a drastic change in the propagation direction of the solitons.
After the collision, the angle between the collision 
direction and the $z$ axis of polarization
 changes to the final value $\theta_f=0$ from the initial value $90\degree$ before collision. 
A video clip of the inelastic collision process of Fig. \ref{fig11} is available
in supplementary  file  change-direction-propagation.mp4  and also available at 
 {\color{blue} https://youtu.be/IIH1y7zrfpk} .

\begin{figure}[!t]
\begin{center}
  \includegraphics[width=\linewidth]{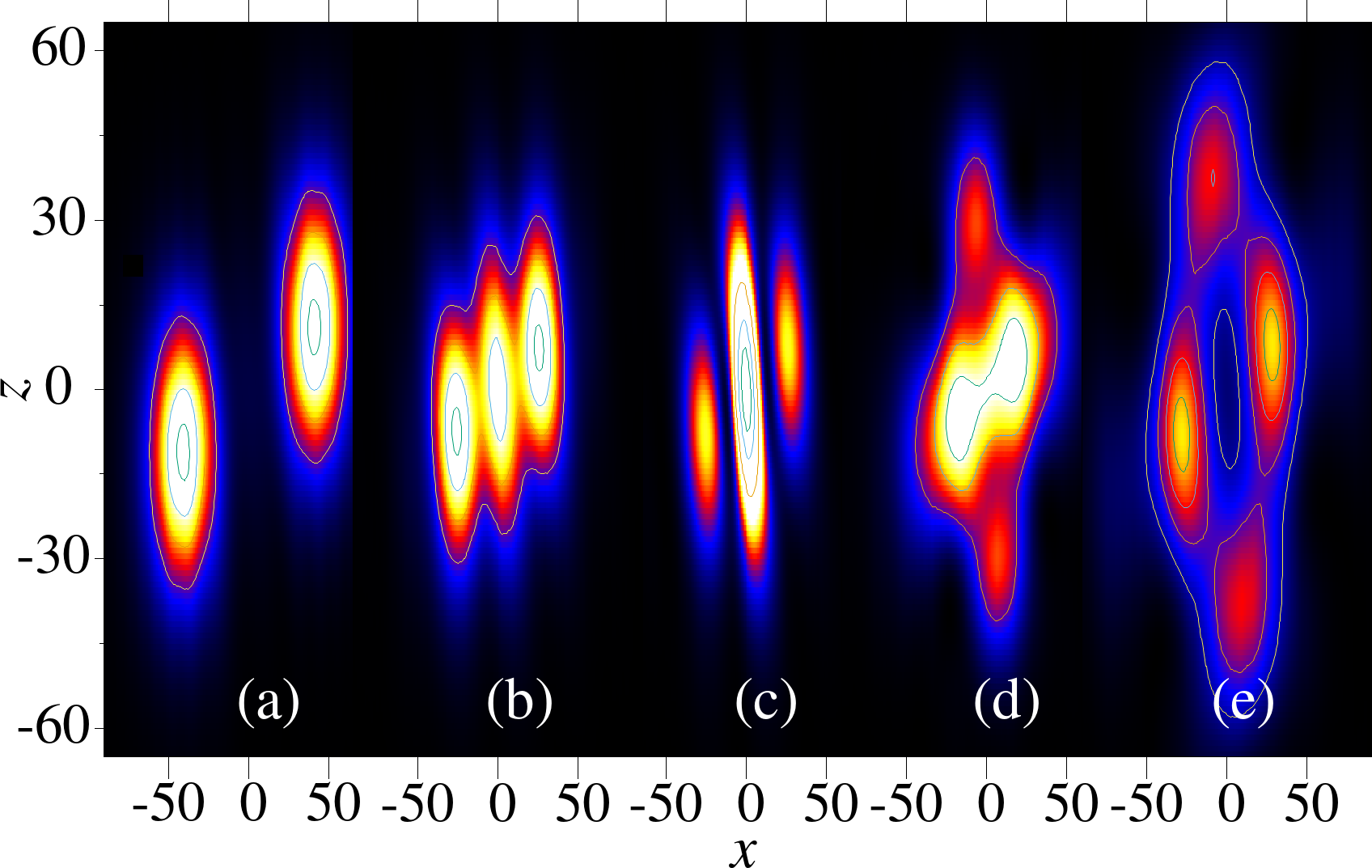}
\caption{
Contour plot of densities $|\phi(x,z)|^2$ 
in a collision with initial $\theta=75\degree$
and velocity $v_0=0.07$ at different times  before (a) $t=100$, during (b) $t=400$ and after (c)  $t=600$, (d) $t=1000$ and (e) $t=1200$ collision.
The initially independent solitons become bound into {an excited  coalesced state,  which eventually decays into four pieces, resembling a quadrupole excitation.}  (same color bar as in Fig. \ref{fig11}).
}
\label{fig12} 
\end{center}
\end{figure}

\subsubsection{Excitation to a quadrupole mode}

\label{4.2.3}

We show in Fig. \ref{fig12}  the evidence of coupling to a quadrupole  mode, through a plot of  the contour of density $|\phi(x,z)|^2$ 
of two anisotropic solitons with $g=8,$ $a_{\mathrm{dd}}/a=3.35$ each in a collision with initial velocity $v_0=0.07$
and $|\theta|=75\degree$ for different times
before (a) $t=100$, during (b) $t=400$ and after (c)  $t=600$, (d) $t=1000$ and (e) $t=1200$ collision.
This velocity corresponds to region (G) of Fig. \ref{fig5}.
Independently of the value of $\theta$, the solitons form a highly excited coalesced  soliton after collision, which eventually decays into four pieces. {The formation of the coalesced soliton is complete in (c). In plots (b) and (c) the density distribution is similar to dipole excitation,  
then  it enters a configuration in (d) resembling a quadrupole excitation, then expands further in (e) maintaining the quadrupole shape. It continues this expansion and breaks up into four pieces, which move apart and never come back} 
{due to an excess of energy of the coalesced soliton. This process is found in all initial angles $\theta$ for $v\approx 0.07$. }
A video clip of this deep inelastic dynamics of Fig. \ref{fig12}
is  contained in supplementary
file    quadrupole-mode.mp4 and also available at 
{\color{blue}  {\color{blue} https://youtu.be/EoaWDzmU-qE} .}

\section{Summary}
\label{sec5}

To summarize, using a numerical solution of the  mean-field 3D and quasi-2D  GP equations, we studied the formation of  {anisotropic} quasi-2D  dipolar BEC solitons \cite{2dsol2}, mobile in the $x-z$ plane,  polarized 
along the axial $z$ direction and harmonically trapped along the transverse $y$ direction. 
In this paper { we considered frontal collision as well as  collision with a non-zero impact parameter
between two anisotropic dipolar soliton  at different velocities}  and for 
different angles between the direction  of motion  and the polarization direction. At large velocities of collision, the collision is found to be quasi elastic with no visible deformation of soliton profiles after collision. {As the velocity is reduced,  the collision becomes inelastic with deformation of soliton profiles after collision and the deformation increases at slower velocities. For even slower velocities the solitons form an unstable coalesced soliton in a highly excited state which decays eventually into more than two pieces.  In this region we identified  a quadrupole excitation of the coalesced soliton and a subsequent decay to four solitons. For further reduction of velocity, the coalesced soliton decays into two solitons moving away in direction different from the incident direction.  }
At sufficiently small velocities  the  fused coalesced soliton does not have  sufficient energy for break up
and performs oscillation in complex modes {(soliton fusion)}; the signature of scissors mode (angular oscillation) and breathing mode (radial expansion and contraction) are found 
in  numerical simulation. 
{  The result of collision for frontal collision and collision with a non-zero impact parameter 
 are quite similar for both  quasi-elastic, viz. Sec. \ref{4.2},  and inelastic collision (not reported here). It has been demonstrated numerically that these robust anisotropic quasi-2D dipolar solitons can be created in a laboratory from trapped dipolar BECs by suitably changing the trap frequency and the scattering length thus making the collision  of such solitons possible in a laboratory \cite{malomed}.     }

\section*{Credit author statement}

Both authors were responsible for Conceptualization, Methodology, and Software. L. E. Young-S. was responsible for Formal Analysis, Investigation, Data Curation and Writing - Original Draft. S. K. Adhikari was responsible for Writing - Review and Editing and Supervision.

 \section*{Acknowledgments}
We thank Prof. L. Santos for his kind interest in this study and many valuable suggestions.
S.K.A. acknowledges support by the CNPq (Brazil) grant 301324/2019-0, and by the ICTP-SAIFR-FAPESP (Brazil) grant 2016/01343-7.

\end{document}